\documentclass[aps,prd,superscriptaddress,notitlepage,nofootinbib,reprint,noeprint]{revtex4-2}
\usepackage{amsmath,amssymb,amsfonts,amsbsy}
\usepackage{xcolor}
\usepackage{graphicx}
\usepackage{multirow}
\usepackage[normalem]{ulem} 
\usepackage{enumerate}
\usepackage{footmisc}

\begin{document}
\title{Confining modeling of quark propagator}

\author{A.E. Radzhabov}\email{aradzh@icc.ru}
\affiliation{Matrosov Institute for System Dynamics and Control Theory SB RAS, 664033, Irkutsk, Russia}

\author{X.L. Shang}\email{shangxinle@impcas.ac.cn (corresponding author)}
\affiliation{Institute of Modern Physics, Chinese Academy of Sciences, Lanzhou 730000, China}

\begin{abstract}

A confining extension of the quark model with nonlocal currents is proposed.
The quark propagator is modified by introducing a cut in $\alpha$-space, which in momentum space corresponds to the subtraction of pole singularities.
A two-phase phase structure is proposed for modeling the confinement-deconfinement phase transition.
In the confined phase, the quark propagator does not have any pole singularities, while in the deconfined phase, there is a single quark pole.
\end{abstract}

\maketitle
\section{Introduction}

Understanding of strong interactions is a complex task.
The fundamental theory of strong interactions, quantum chromodynamics (QCD), is formulated in terms of quarks and gluons.
Experimentally observed hadrons are interpreted as bound states of these fundamental constituents. This is due to the nonperturbative nature of QCD. The strong coupling constant $\alpha_s$ is a small parameter only at large $Q^2$.
Free quarks and gluons have not been observed in experiments, they are confined inside hadrons.
It is expected that under certain physical conditions, such as in heavy ion collisions or within neutron stars, a different state of matter may form, which differs from the hadronic world we know. In this state, fundamental particles become deconfined and can freely propagate as quarks and gluons in a quark-gluon plasma (QGP).
The only ab-initio, nonperturbative method for investigating strong interactions is lattice QCD simulation. In recent years, lattice QCD has made impressive progress.
Despite this, the theoretical understanding of the underlying degrees of freedom still requires modeling since lattice simulations only provide numerical results without deeper insight.
The symmetries of strong interactions can serve as a guiding principle, such as chiral perturbation theory \cite{Gasser:1983yg}, which deals with Goldstone bosons or the quark version of Nambu-Jona-Lasinio (NJL) model \cite{Nambu:1961fr}, where chiral symmetry is spontaneously broken due to quark condensate and massive constituent quarks are formed instead of current ones.  

More sophisticated approaches are based on coupled systems of Dyson–Schwinger equations (DSE) for quark and gluon
propagators and their vertices, along with Bethe–Salpeter equations for hadronic bound states \cite{Roberts:2000aa}. 
These approaches describe quarks with a momentum-dependent mass  $m(k^2)$ and wave-function renormalization  $Z(k^2)$. 
At large momentum transfer, the mass function tends to the current mass. 
The nonlocal version of the Nambu-Jona-Lasinio (NJL) model \cite{GomezDumm:2001fz, Dorokhov:2000gu} is somewhere between these two approaches. The structure of the interaction is similar to that of the NJL model, but the nonlocal quark current leads to a quark propagator with momentum-dependent mass.
 
However, the absence of confinement can make calculations problematic. %
If the denominator of the quark propagator has zeroes\footnote{This expression is given in the Euclidean metric,  which is mainly used in paper. Only equations with gamma matrices \eqref{QuarkProp} and \eqref{VirtonModel} are given in Minkowski momentum space.}
\begin{align}
&{k^2+m^2(k^2)}=0, \label{DenomProp}
\end{align} 
at real values, $k^2=-m^2_{\mathrm{pole}}$,
this will result in an imaginary part in the polarization operators of mesons.
This imaginary part corresponds to the appearance of free quarks.
One way to avoid the occurrence of an imaginary component in polarization operators is to have a mass function $m(k^2)$ that only results in complex-valued solutions of this equation.
This case is similar to the %
scheme \cite{Cutkosky:1969fq} and could be interpreted as modeling confinement \cite{Bhagwat:2002tx}.
Apart from technical issues in calculating Feynman diagrams, the physical meaning of complex singularities remains unclear.
In a medium with varying temperature or baryon density, the positions of the singularities change, further complicating their physical interpretation.

In this paper, we aim to demonstrate that by making relatively simple modifications to the model, it is possible to eliminate these undesirable singularities.
The general idea is that confinement should somehow change the model, which is based on chiral symmetry.
Therefore, physical observables related to free quarks should not exist in the hadron phase.
In addition, a method for realizing the confinement-deconfinement phase transition is required.
It should be noted that the Polyakov loop extension of the NJL model with an effective potential of gauge degrees of freedom is intended to address one aspect of quark confinement at finite temperature -- the suppression of quark pressure in the confining phase \cite{Ratti:2005jh}. 
We propose an additional extension to the quark model.
The inverse Laplace transform of the denominator of the quark propagator in the confined phase is modified to ensure that the transformed function is valid for arbitrary momentum.
The resulting momentum space-transformed function has no pole singularities.
The new momentum scale parameter associated with this modification can be interpreted as the confinement scale $\Lambda_c$.
In the deconfined phase, a similar procedure is applied when the quark pole has been separated.

\section{Nonlocal model}
\label{model}
The Lagrangian of the $SU(2)$ nonlocal chiral quark model with pseudoscalar–scalar sectors is:  
\begin{align}
&\mathcal{L}= \mathcal{L}_{free}+\mathcal{L}_{P,S} %
 ,\quad %
\mathcal{L}_{free} = \bar{q}(x)(i \hat{\partial}-M_c)q(x), \\
&%
\mathcal{L}_{P,S} = \frac{G}{2}\bigg(\Big(J_S^a(x)\Big)^2+\Big(J^a_P(x)\Big)^2\bigg) ,\quad %
\end{align}
where $M_c=\mathrm{diag}(m_c,m_c)$ is the current quark mass matrix with diagonal elements $m_c$, $G$ is %
the four-quark coupling constant. %

The nonlocal quark current can be taken in one-gluon-exchange-like (OGE) \cite{GomezDumm:2001fz} or instanton liquid model (ILM) \cite{Plant:1997jr,Dorokhov:2000gu} forms. The structure of OGE-type currents is:
\begin{align}
J_{M}^{a\{,\mu\}}(x)
= \int d^4x_1 d^4x_2 \,
\delta\left(x-\frac{x_1+x_2}{2}\right) \times \notag \\
\times g((x_1-x_2)^2)\,
\bar{q}(x_1) \, \Gamma_{M}^{a\{,\mu\}} \, q(x_2), \label{eq2} %
\end{align}
with $M=S,P$ and  %
$\Gamma_{{S}}^{a}=\lambda^{a}$, $\Gamma_{{P}}^{a}=i\gamma^{5}\lambda^{a}$. %
For the $SU(2)$ model,  
the flavour matrices are Pauli matrices: $\lambda^{a}\equiv\tau^{a}$, $a=0,..,3$ with $\tau^0=1$.
$g(x)$ is the form factor encoding the nonlocality of the QCD vacuum. 

The bosonized Lagrangian after the Hubbard-Stratonovich transformation is: 
\begin{align}
\mathcal{L}_{eff}&=
\bar{q}(x)(i \hat{\partial}_x -M_c)q(x)+\sigma_0 J_S^0(x) +\notag \\
& - \frac{1}{2 G}\left(\Big(P^a(x)\Big)^2+ \Big(\tilde{S}^a(x)+ \sigma_0 \delta^a_0\Big)^2\right)+\label{Bosonized}\\
&%
+
P^a(x)J_{P}^{a}(x)+
\tilde{S}^{a}(x)J_{S}^{a}(x) \notag%
\end{align}
Spontaneous chiral symmetry breaking leads to non-zero vacuum expectation value of the scalar isoscalar field  
$\langle S^0 \rangle_0=\sigma_0\neq0$.
After shifting the scalar isoscalar field $S^0=\tilde{S}^0+\sigma^0$,
the momentum-dependent
quark mass appears\footnote{The same notation is used for the Fourier transform of functions.}:
\begin{align}
m(p^2)=m_c+m_{d}g(p^2) \label{Mass}
\end{align}
where scalar coefficient $m_{d}=-\sigma^0$
can be found from self-consistent equation
\begin{align}
m_{d}= G \frac{8  N_c }{(2 \pi)^4} \int d^4_Ek
\frac{g(k^2)m(k^2)}{k^2+m^2(k^2)}, \label{Gap}
\end{align}
where $N_c$ is the number of quark colors.
The quark propagator in Minkowski metric is 
\begin{align}
\mathrm{S}(p)=(\hat{p}-m(p))^{-1}. \label{QuarkProp}
\end{align}

The pion polarization loop is
\begin{align}
&\Pi_{\pi}(p^2)=\frac{8 N_c}{(2\pi)^4} \int d_E^4 k f^2(q^2)\, \mathrm{D}(k_-^2)\mathrm{D}(k_+^2) \times \notag \\
&\times \left[ 
{(k_- \cdot k_+)+m(k_+^2)m(k_-^2)} \right]
,
\label{poloperOGE} %
\end{align}
where %
$\mathrm{D}(k^2)=(k^2+m^2(k^2))^{-1}$ is scalar propagator %
and momenta are $p=k_+-k_-$ and $q=(k_++k_-)/2$. 
In the general case, the flow of momentum in the diagram is arbitrary, with  $k_+=k+\zeta p$ and $k_-=k-(1-\zeta) p$, where $0<\zeta<1$.
The result should be independent of this choice. 
We utilize this feature to verify calculations.
The pion mass $M_\pi$ can be found from equation\footnote{For numerical estimation the set of model parameters with $\bar{q}q=-(240 $MeV$)^3$ \cite{GomezDumm:2006vz}(Scheme II) is used: $m_c=5.8$ MeV, $m_d=424$ MeV, $\Lambda=752.2$ MeV.}
\begin{eqnarray}
-G^{-1}+\Pi_\pi\left(-M_\pi^2\right)=0.
\end{eqnarray}
%
%
%
%
%
%
%
%
%
%
%
%
%

\subsection{Singularities of quark propagator}

\begin{figure}[t]
    \centering
        \includegraphics[width=0.45\textwidth]{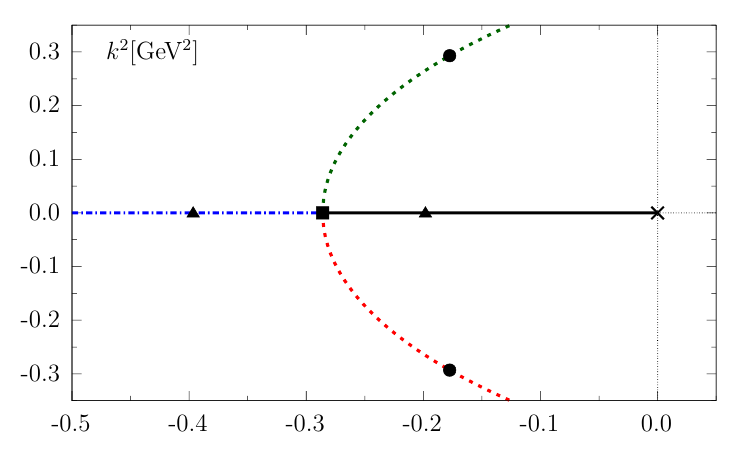}
\caption{%
The position of the poles of the quark propagator in the complex $k^2$ plane (in GeV$^2$) is shown. Lines represent the position of the pole with varying $m_d$: black solid line is the first real-valued solution for $m_d<m_d^{crit}$, blue dashed-dotted line is second real-value solution for $m_d<m_{d}^{crit}$, and red and green dotted lines represent complex conjugated solutions for $m_d>m_{d}^{crit}$. The symbols indicate the following solutions: circles represent  vacuum value of $m_d$, square represent solution for $m_d^{crit}$, triangles represent solutions with the property $z_2 - z_1 = z_1$, which is used as a confinement-deconfinement criterion, and cross represent zero $m_d$.
}
\label{CompPolPosition}
\end{figure}

Analytical structure of quark propagator in nonlocal model strongly depends on the form-factor. For a Gaussian form-factor
\begin{align}
g(k^2)=\exp(-k^2/\Lambda^2), \label{GForm}
\end{align} 
the denominator of the quark propagator \eqref{DenomProp} has an infinite number of solutions  for complex values of $k^2$.
The inverse function, which is a %
scalar propagator $\mathrm{D}(k^2)$, has corresponding pole singularities.
%
The structure of the quark propagator is similar to that found in DSE studies \cite{Alkofer:2003jj,Dorkin:2014lxa,Windisch:2016iud}, where complex-valued singularities have also been observed.
For some model parameters, the first two poles could be for real negative-valued $k^2$, that is, $k^2 = -m^2_\mathrm{pole}$. After analytically continuing the meson polarization loop \eqref{poloperOGE} to Minkowski space, $p^2$ becomes negative. The imaginary part appears when  $-p^2$ exceeds the threshold mass squared, $\mathrm{M}_{\mathrm{thr}}^2 = 4m_\mathrm{pole}^2$.
For complex-valued poles the situation is %
more complicated. 
Imaginary parts of different poles cancel each other \cite{Bhagwat:2002tx} but the real part of the polarization loop has a cusp\footnote{%
Here $\mathrm{M}_{\mathrm{thr}}^2=2\mathrm{Re}(m^2_{\mathrm{pole}}) + 2 \sqrt{\mathrm{Re}(m^2_{\mathrm{pole}})^2+\mathrm{Im}(m^2_{\mathrm{pole}})^2}$.}. Such a feature would not seem to be a physical one, as it would be observable through experiment. 
In order to perform calculations in a medium, the $G$ and $\Lambda$ are treated as constants and $m_d$ is the only parameter that changes in the quark sector. Therefore, it is interesting to investigate the position of the poles of the quark propagator for an arbitrary value of %
$m_d$ (see Fig.~\ref{CompPolPosition}), since this is exactly the scenario that would occur in a medium.
At the vacuum value of $m_d$, the first two poles are complex-valued (circles in Fig.~\ref{CompPolPosition}). 
As $m_d$ decreases, the two poles start to move towards the real axis along the dashed lines.
At $m_d^{crit}$, these complex-conjugate poles become real-valued (square).
At finite temperature, %
$m_d^{crit}$ corresponds to chiral phase transition point.
If the presence of real-valued poles is considered an indicator of the confinement-deconfinement phase transition, these two phase transitions synchronize.

However, %
the problems are:
\begin{enumerate}
\item 
How to perform analytical continuation of the meson polarization loops integral beyond point $\mathrm{M}_{\mathrm{thr}}$? %
The prescription for complex-valued poles \cite{Cutkosky:1969fq,Bhagwat:2002tx} leads to the suppression of the imaginary part of polarization loops. However, there is a cusp in the real part of the loop at the pinch point,  %
which seems unphysical. %
\item 
What is the physical meaning of the second real-valued solution? 
It has a wrong sign of residue, which make it similar to Lee-Wick QED model with Pauli-Villars regulator \cite{Lee:1969fy}.
The first real solution with decreasing of $m_d$ goes to zero, meaning the quark becomes a current one, while the second mass in this limit approaches infinity. At $m_d^{crit}$ they have the same value.
\end{enumerate}

\subsection{Suggestions for remove quark thresholds}
Two directions of confinement modeling in effective approaches can be noted. The first approach is to represent the quark propagator as an entire function whose parameters are determined by fitting to observables:
\begin{enumerate}
\item In {\it virton} model \cite{Dubnickova:1979mq} the quark propagator is (in Minkowski metric)
\begin{align}
\mathrm{S}(p)=-M\exp\biggl(-l\hat{p}-p^2 L^2/4\biggr), \label{VirtonModel}
\end{align}
where $M$, $l$ and $L$ are some model parameters.
\item In DSE studies \cite{Burden:1995ve, Hecht:2000xa,Chen:2021guo} of hadron properties the 
scalar and vector part of dressed quark propagator $\mathrm{S}(p)=-i\hat{p}\sigma_V(p^2)+ \sigma_S(p^2)$ (in Euclidean space) is expressed in the form 
\begin{align}
\sigma_S(x) & = \lambda \bigl(2\,\bar m \,{\mathcal F}(2 (x+\bar m^2)) +\notag \\
& \quad
 + {\mathcal F}(b_1 x) \,{\mathcal F}(b_3 x) \,
\left[b_0 + b_2 {\cal F}(\epsilon x)\right]\bigr),\notag \\
\sigma_V(x) & = \lambda^2 \frac{1 - {\cal F}(2
(x+\bar m^2))}{x+\bar m^2}\, , 
\end{align} 
where 
\begin{align}
&\mathcal{F}(x) = \frac{1 - \mathrm{e}^{-x}}{x} 
\end{align} 
and dimensionless values are $x=p^2/\lambda^2$, $\bar m$ = $m/\lambda$. The mass-scale, $\lambda$, and model parameters $m$, $b_0$, $b_1$, $b_2$, $b_3$, $\epsilon$ are fitted in analyses of light-meson observables \cite{Burden:1995ve, Hecht:2000xa}.
\item Scalar (or vector) part of quark propagator as entire function in nonlocal model \cite{Radzhabov:2003hy}
\begin{align}
\mathrm{D}(k^2)=\mathcal{F}({\left(k^2+m_c^2\right)}/{\Lambda^2}). 
\end{align} 
\end{enumerate}
The other direction involves changing the method of calculating loop integrals:
\begin{enumerate}
\item In the quark confinement model of hadrons
\cite{Efimov:1993zg}, the confinement is realized through ansätze for quark propagator averaged in gluon background. 
\item IR-cutoff in NJL model with proper time regularization \cite{Ebert:1996vx} 
with function
\begin{align}
&\frac{1}{s+M^2}=\int\limits_{0}^{\infty} d\tau \mathrm{e}^{-\tau(s+M^2)} 
\quad \to \label{EbertQQ} \\
&\to
\int\limits_{\tau_{UV}^2}^{\tau_{IR}^2} d\tau \mathrm{e}^{-\tau(s+M^2)} = \frac{\mathrm{e}^{-\tau_{UV}(s+M^2)} -\mathrm{e}^{-\tau_{IR}(s+M^2)}}{s+M^2}\notag
\end{align}
or with three dimensional cut-off \cite{Blaschke:2001yj}.
The function \eqref{EbertQQ} is quite often used to model confined propagators \cite{Gutierrez-Guerrero:2010waf,Roberts:2010rn,Wang:2013wk,Marquez:2015bca}.

\item 
Cut in $\alpha$-space \cite{Branz:2009cd} of the whole expression for polarization loops. %
The expression for the polarization loop with %
quark propagators is represented by the usual prescriptions in $\alpha$-space (Schwinger parameterization). In the integral over the sum of $\alpha$, the upper limit is changed from infinity to $1/\lambda^2$. %

\end{enumerate} 

%
%
%
%
In a nonlocal model, the problem with a momentum-dependent mass $m(p^2)$ is that for gauge invariance, it is necessary to modify the vertices of interaction with external currents, along with the quark propagator. As a result, it is unclear how to implement the modification of the loop integral.

\subsection{Confining $\alpha$-prescription for quark propagator}%
If the %
scalar propagator has only pole singularities, it can be represented as the sum of its poles
\begin{align}
&\mathrm{D}(k^2)= \sum\limits_{i=1}^{N} \frac{R_{z_i} }{k^2+z_i} \label{SeriesMomentum1} 
\end{align} 
where $N$ is number of poles (for Gaussian form-factor $N=\infty$%
). %
Poles are sorted by the size of the real part of  $z_i$ with  $z_1$  having the smallest real part. 
   
Another expansion of the $\mathrm{D}(k^2)$ %
is based on its high-energy $k^2$
behavior, i.e., expansion over $m_d$
\begin{align}
&\mathrm{D}(k^2)= \frac{1}{k^2+m_c^2}-\frac{2 m_c m_d g(k^2) }{(k^2 + m_c^2)^2} %
+O(m_d^2).  \label{SeriesMomentum2} 
\end{align} 

The Laplace transform is defined as:
\begin{align}
&\mathrm{D}(k^2)= L \left\{\mathbf{D}(\alpha)\right\},
\quad
\mathrm{D}(k^2)=%
\int\limits_{0}^{\infty} d\alpha \mathrm{e}^{-\alpha k^2} \mathbf{D}(\alpha). \label{LaplaceTransform}
\end{align} 
Series corresponding to the above equations \eqref{SeriesMomentum1} and \eqref{SeriesMomentum2} can be obtained with inverse Laplace transforms  $\mathbf{D}(\alpha)= L^{-1} \left\{\mathrm{D}(k^2)\right\}$ for each term:
\begin{align}
\mathbf{D}(\alpha)&=\sum\limits_{i=1}^{\infty} {R_{z_i}} \mathrm{e}^{-\alpha z_i}  ,  \label{SeriesLaplace1} \\
\mathbf{D}(\alpha)&= \mathrm{e}^{-\alpha m_c^2} -2m_c m_d \mathrm{e}^{-\alpha_1 m_c^2} \alpha_1 \theta(\alpha_1) +.. , \label{SeriesLaplace2} 
\end{align} 
where $\alpha_i=\alpha-1/\Lambda^{2i}$. The first series is valid for arbitrary form-factors. 
In the second one, only the first term is model-independent, and the actual analytical form of the form factor \eqref{GForm} should be used for the other terms%
\footnote{
The inverse Laplace transform of the terms with $g(k^2)$ \eqref{GForm} is
\begin{align}
 L^{-1} \left\{\frac{g^n(k^2) }{(k^2 + m_c^2)^m} \right\}=\mathrm{e}^{-\alpha_n m_c^2}\frac{(\alpha_n)^{m-1} }{(m-1)!}
  \theta(\alpha_n).
\end{align}}.
These expansions complement each other, because in order to obtain $\mathbf{D}(0)=1$, an infinite number of terms in series \eqref{SeriesLaplace1} must be summed. In contrast, the first term in series \eqref{SeriesLaplace2} immediately gives the correct answer. On the other hand, in order to obtain the behavior of $\mathbf{D}(\alpha)$ for large $\alpha$, the first terms are sufficient in \eqref{SeriesLaplace1}.

\begin{figure}[t]
    \centering
        \includegraphics[width=0.45\textwidth]{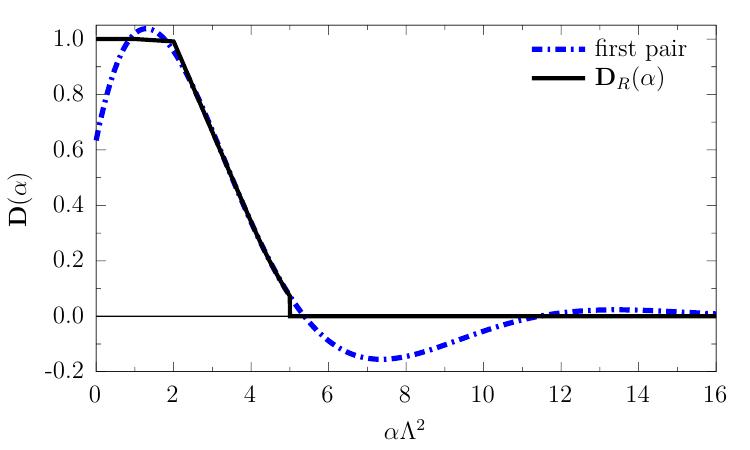}
\caption{Alpha transformed function $\mathbf{D}(\alpha)$ of %
scalar propagator: %
blue dash-dotted line is contribution of first pair of complex-conjugated poles in \eqref{SeriesLaplace1} and solid line is the $\mathbf{D}_R(\alpha)$ from \eqref{InvDRalpha} with $\Lambda_c^2= \Lambda^2/{5}$.%
}
\label{LapImag}
\end{figure}

The region of convergence for the Laplace transform \eqref{LaplaceTransform} is $k^2 > -Re(z_1)$.
In order to make this transformation applicable to arbitrary $k^2$, it is necessary to change the behavior of the function $\mathbf{D}(\alpha)$ for large values of $\alpha$.
The simplest modification is following\footnote{%
At first glance, the only physical restriction seems to be  $\Lambda_c < \Lambda$. %
Otherwise, the only first term in the series \eqref{SeriesLaplace2} would give a nonzero contribution, and 
$\mathbf{D}_R(\alpha)$
becomes  $m_d$-independent.
\label{FootnoteLambdaC1}} %
\begin{align}
\mathbf{D}_R(\alpha)=\mathbf{D}(\alpha)\theta\bigl({1}/{\Lambda_c^2}-\alpha\bigr)%
\label{InvDRalpha}
\end{align} 
In this case, the Laplace transformation \eqref{LaplaceTransform} becomes valid for arbitrary $k^2$.

Performing Laplace transform, one can obtain the modified scalar %
propagator in momentum space %
\begin{align}
\mathrm{D}_R(k^2)&=\frac{1}{k^2+m_R^2(k^2)}=\notag\\%
&=\mathrm{D}(k^2)
- \sum\limits_{i=1}^{\infty} \frac{R_{z_i} Q\left(k^2+z_i\right)}{k^2+z_i},\label{InvDR} 
\end{align} 
where $Q(k^2)=\exp\left(-k^2/\Lambda_c^2\right)$. $\mathrm{D}_R(k^2)$ is entire function, since $Q(0)=1$ and all the singularities have been subtracted. On the other hand, as $k^2$ becomes large, $\mathrm{D}_R(k^2)$ tends to  $\mathrm{D}(k^2)$, since $\Lambda_c< \Lambda$.

From the equation \eqref{InvDR}, one can obtain the mass function\footnote{
Here, one may encounter a problem that for a real $k^2$, the expression under the square root becomes negative, i.e., the quark mass $m_R(k^2)$ becomes imaginary. 
In order for $m_R(k^2)$ to be real, it is necessary that the  $\mathrm{D}_R(k^2)$ function be non-negative. The behavior of the  $\mathrm{D}_R(k^2)$ function at large negative $k^2$ 
for nonzero $\mathbf{D}(1/\Lambda_c^2)$ is  
\begin{align}
\mathrm{D}_R(k^2) \approx \frac{\mathrm{e}^{-k^2/\Lambda_c^2}}{-k^2}\mathbf{D}(1/\Lambda_c^2).
\end{align}
Therefore, the condition $\mathbf{D}(1/\Lambda_c^2) \geq 0$ must be satisfied, which approximately implies that $\Lambda_c^2\geqslant \Lambda^2/{5.4}$, as shown in Fig.\ref{LapImag}.

%
\label{FootnoteLambdaC2}
} $m_R(k^2) = \sqrt{\mathrm{D}_R^{-1}(k^2) - k^2}$. 

The behavior of $\mathbf{D}_R(\alpha)$ is shown in Fig.\ref{LapImag}. %
The asymptotic behavior of the function $\mathbf{D}(\alpha)$ is determined by its two lowest complex conjugate poles, i.e. $i=1,2$ in \eqref{SeriesLaplace1}.
As illustrated in the Fig.\ref{LapImag} that this mode occurs at $\alpha$ of the order of $3/\Lambda^2$.

\begin{figure}[t]
    \centering
        \includegraphics[width=0.45\textwidth]{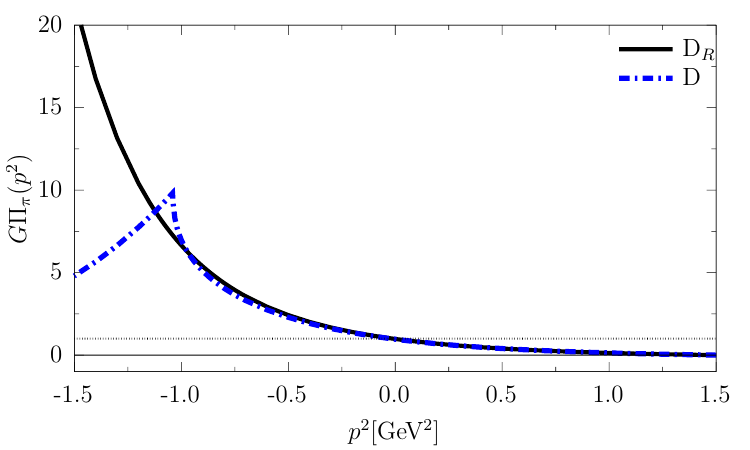}
\caption{Pion polarization loop times four-quark coupling constant in nonlocal model (blue dash-dotted line) and with confining prescription (black solid line). Thin dashed horizontal line denote $1$ which is correspond to meson pole. 
}
\label{PolOperator}
\end{figure}

To calculate the polarization loop in \eqref{poloperOGE}, the prescription for propagators with complex masses can be used \cite{Cutkosky:1969fq}. The main difference is that, instead of pole singularities, branch cut points will appear in the complex plane.
The result of the calculation of the pion polarization loop is presented in Fig. \ref{PolOperator}. One can observe that the cusp-like behavior of the meson polarization loop around $-1$ GeV$^{2}$ is absent for the  $\mathrm{D}_{R}(k^{2})$.
The possible choice of $\Lambda_c$ is as follows: as mentioned in footnote \ref{FootnoteLambdaC1}, $\Lambda_c < \Lambda$.
%
A value of $\Lambda_c = \Lambda / \sqrt{5}$ appears to be a reasonable compromise which leads to a reasonable behavior of $m_R(k^2)$, see footnote  \ref{FootnoteLambdaC2}. This corresponds to a numerical value of $\Lambda_c =336$  MeV.

Since quarks can freely propagate in the deconfined phase, a pole for real negative $k^2$ should be present.
As the poles on the right-hand side of Eq. \eqref{InvDR} occur simultaneously in $\mathrm{D}(k^2)$ and in the second term with sum, it is possible to start the summation at $i=2$.
In this case, %
one can rewrite the expression Eqs.\eqref{InvDR} for $\mathrm{D}_R(k^2)$ in the form 
\begin{align}
\mathrm{D}_R(k^2)=
\mathrm{D}(k^2) - \sum\limits_{i=2}^{\infty} \frac{R_{z_i} 
Q\left(k^2+z_i\right)
 }{k^2+z_i}=\notag\\
\quad=\frac{R_{z_1}}{k^2+z_1-i\epsilon}+\bigg[%
\mathrm{D}(k^2)-\frac{R_{z_1}}{k^2+z_1}- 
\notag \\
-\sum\limits_{i=2}^{\infty} \frac{R_{z_i} 
Q\left(k^2+z_i\right)
}{k^2+z_i}\bigg],\label{InvDRdecon} 
\end{align} 
where the only pole is isolated and the expression in square brackets is an entire function. 
In the limit of $m_d$ approaching zero, the first pole approaches $z_1\to m_c^2$, with a residue of $R_{z_1}  \to 1$ , and the expression within the brackets approaches zero.

The combined two-phase confinement-deconfinement model is:
\begin{enumerate}
\item Confined phase: the subtraction sum in Eq.%
\eqref{InvDR} starts from $i=1$. $\mathrm{D}_R(k^2)$ is an entire function.
\item Deconfined phase: the subtraction sum starts from $i=2$. $\mathrm{D}_R(k^2)$ has one pole which corresponds to the pole quark mass. %
\end{enumerate}
An important question is about symmetries - what will happen to the gauge symmetry and chiral symmetry?
The situation is clearer with gauge symmetry: one simply needs to include $m_R(k^2)$ in the expressions for effective vertices \cite{GomezDumm:2006vz} instead of $m(k^2)$, and the Ward identity is automatically satisfied in this case. 

For chiral symmetry, the situation is more complicated. By construction there is a fine-tuning between the gap equation \eqref{Gap} and the pion polarization loop \eqref{poloperOGE} \cite{Osipov:2007zz} and changing of only the mass due to confinement prescription breaks this fine-tuning.
%
%
The possible solution is to adjust the normalization of $g(p)$ at the pion vertex. 
Numerically, these corrections are around two percent for the model parameters used. 

\section{Finite T behavior}

The mean field thermodynamic potential is %
\cite{Radzhabov:2010dd,Carlomagno:2019yvi}
\begin{align}
\Omega^{MF} =\frac{m_d^2}{2 G} -4 \sum_{i=0,\pm}\int_{k,n}\ln\bigg[k_{n,i}^2+m^2(k_{n,i}^2)\bigg] + \notag \\
+ U (\Phi,\bar{\Phi}) + \Omega_0, \label{ThermodynamicPoten}
\end{align}
where $k_{n,i}=(\omega_n^i-i {\mu})^2+\mathbf{k}^2$ and $\int_{k,n} \equiv T\sum_n\frac{d^3p}{(2\pi)^3}$. Fermionic Matsubara frequencies $\omega_n^i$ are partially shifted due to the presence of a Polyakov loop $\omega_n^{\pm}=\omega_n\pm \phi_3$, $\omega_n^0=\omega_n$ and $\omega_n=(2n+1)\pi T$.
The thermodynamic potential is divergent due to the current quark contribution.
The infinite normalization constant, $\Omega_0$, is chosen such that the pressure should be zero under vacuum conditions, which means at zero temperature, zero chemical potential, and for the vacuum $m_d$ value.

In the presence of the Polyakov loop, the equations of motion for the mean fields are
\begin{align}
\frac{\partial \Omega^{MF}}{\partial m_d} = 0 , \quad \frac{\partial \Omega^{MF}}{\partial \phi_3} = 0.
\label{GapEqAtT}
\end{align}

For comparison, the three models are considered:
\begin{enumerate}[(I)]
\item %
Nonlocal quark model
\item Confined quark model (the sum in eq. \eqref{InvDR} starts from $i=1$)
\item Combined confinement-deconfinement %
model (the sum  in eq. \eqref{InvDR} starts from $i=1$ in confined phase and $i=2$ in deconfined phase.)
\end{enumerate}
The boundary between phases in model (III) is determined by the properties of quark propagator. %
To a first approximation, in the deconfined phase, there are real solutions to Eq. \eqref{DenomProp}, whereas in the confined phase, only complex-valued solutions are possible.
However, it seems that the region where two real poles almost coincide should also be included in the confined phase. In practice, we use a criterion that the difference between the two real poles is of the order of the first one, i.e., in the deconfined phase, $z_2 - z_1 > z_1$. The corresponding points of $m_d$ are approximately 319 MeV, which is the border between real and complex pole solutions, and 310 MeV when $z_2 - z_1 > z_1$ (the position of the poles is shown in Fig.\ref{CompPolPosition}).

\begin{figure}[ht]
    \centering
        \includegraphics[width=0.45\textwidth]{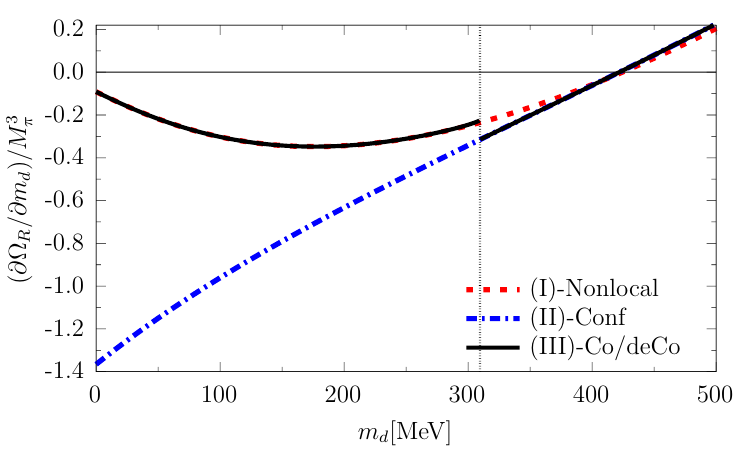}
        \includegraphics[width=0.45\textwidth]{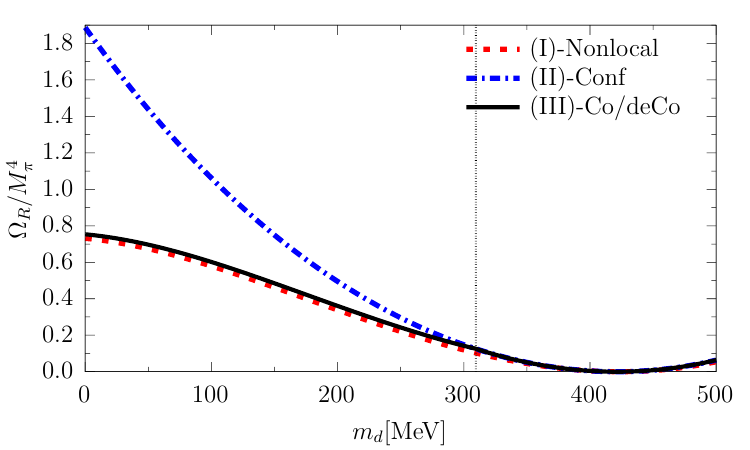}
\caption{${\partial \Omega_R}/{\partial m_d}$ and $\Omega$ in vacuum as a function of $m_d$ for three models: red dotted line nonlocal quark model (I), blue dash-dotted line is for confining model (II) and solid line for model with confinement/deconfinement transition (III). Thin vertical line corresponds to confinement/deconfinement border. }
\label{EffPotVac}
\end{figure}

Since the modification \eqref{InvDR} and \eqref{InvDRdecon} is $m_d$-dependent one should include this behavior when taking the derivatives from the thermodynamic potential \eqref{ThermodynamicPoten} in \eqref{GapEqAtT}.
Therefore if one uses the analytical form of thermodynamic potential eq. \eqref{ThermodynamicPoten}, the expression for the $m_d$-gap equation should be changed. Alternatively, one could keep the analytical form of the partial derivative 
\begin{align}
\frac{\partial \Omega_R}{\partial m_d} =\frac{m_{d}}{G}-8 \sum_{i=0,\pm}\int_{k,n} 
g(k_{n,i}^2)m_R(k_{n,i}^2)\mathrm{D}_R(k_{n,i}^2)\label{GapT},
\end{align}
and calculate the thermodynamic potential numerically 
\begin{align}
\Omega_R = \int \limits_{0}^{m_d} d m_d \frac{\partial \Omega_R}{\partial m_d} + C(T,\mu),
\end{align}
using  $C(T, \mu)$ as the integration constant. $C(T, \mu)$ is equal to the value of $\Omega^{MF}$, taken at zero $m_d$,
for a given values of $T$, $\mu$, and $\phi_3$ with normalization $\Omega_R=0$ in vacuum.
It is very instructive to study the $m_d$ dependence of the thermodynamic potential in almost vacuum, since there is no influence from the Polyakov loop in this case.
In Fig. \ref{EffPotVac}, the behavior of $\Omega_R$ and  ${\partial \Omega_R}/{\partial m_d}$  are shown for models (I)-(III) as a function of $m_d$. %

\begin{figure}[ht]
    \centering
        \includegraphics[width=0.45\textwidth]{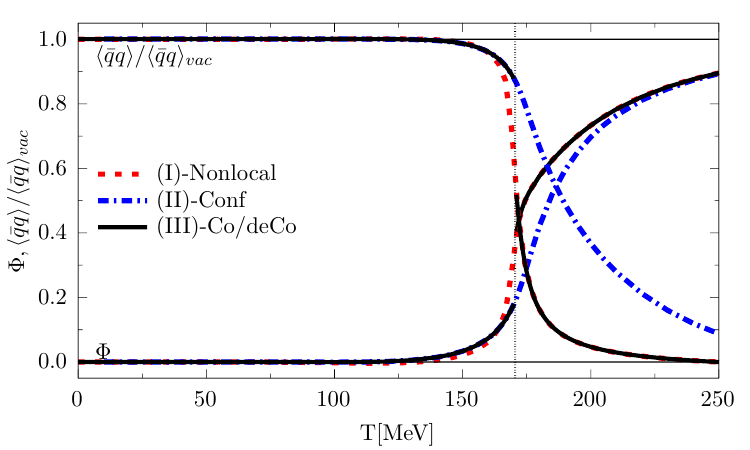}
\caption{Finite temperature behavior of quark condensate divided to vacuum value (upper curves at left side) and Polyakov loop (downer curves at left side) for three models: red dotted line nonlocal quark model (I), blue dash-dotted line is for confining model (II) and solid line for model with confinement/deconfinement transition (III). }
\label{FiniteTCondPolLoop}
\end{figure}
The finite temperature behavior of the quark condensate and the Polyakov loop are shown in Fig. \ref{FiniteTCondPolLoop}. 
One can see here that even for the confining model (II), the quark condensate (and similarly $m_d$) decreases as the temperature increases.
The reason for this is the gap equation \eqref{GapEqAtT}. In the numerator of the integrand \eqref{GapT}, there is a form factor $g(k_{n,i}^2)$. The Matsubara frequency is proportional to $T$, and the first fermionic frequency is nonzero ($\omega_0 = \pi T$). Therefore, as the temperature increases, the second term on the right-hand side of \eqref{GapT} becomes strongly suppressed due to the form factor. Therefore, in order to fulfill the gap equation, a corresponding decrease of $m_d$ is needed.

The finite $T$ phase transition in model (III) is of first order. This is due to the fact that the gap equations in the two phases are not completely synchronized. Fig. \ref{EffPotVac} shows that the gap equation (upper part) is not smooth, and there is a jump in the behavior of ${\partial \Omega_R}/{\partial m_d}$ between the phases. In some conditions, both solutions are possible.
However, as can be seen in Fig. \ref{FiniteTCondPolLoop}, the finite $T$ behavior of model (III) (black solid lines) is not significantly different from that of the model (I) (red dashed lines).

At finite chemical potential, the second minimum at low $m_d$ appears in models (I) and (III), i.e., in the deconfined phase. %
At a certain critical value of $m_d$, the system undergoes a transition from a confined to a deconfined state.
In model (II), this is not the case by design.
\section{Conclusions}

The paper discusses a simple method based on modifying the Laplace transform of the quark propagator to phenomenologically model confinement. The main goal is to improve the analytical behavior of the quark propagator. Two phases are considered at finite $T/\mu$: in the confined phase, the quark propagator has no pole singularities, while in the deconfined phase it has a single physical pole. This leads to mean field calculations for the critical temperature and chemical potential that are almost unchanged after the improvement.
%
%
%
%
%
Pole singularities in the complex plane are absent, and therefore there are no physical consequences that can be caused by poles. Instead, the mass function has cuts in the complex plane, and instabilities discussed in \cite{Benic:2012ec} are present, although they are less significant.

One can expect that a more interesting situation will arise with $1/N_c$ corrections. 
Namely, the mesonic polarization loops at leading order in the nonlocal model exhibit a cusp-like behavior when poles in the complex plane are located inside the integration contour. For finite temperature and zero chemical potential, one can calculate the mesonic correction to the quark condensate using only the Euclidean properties of the polarization loops. At finite chemical potential, the cusp-like behavior of meson polarization loops can make calculations challenging.

However, our prescription is not free from problems. Some of these problems are technical and can be easily solved. In order for the pion to be massless in the chiral symmetric case, when $m_c\to 0$, it is necessary to modify %
the pion-quark vertex. The value of $g(0)$, which represents the coupling between the pion and quark, should slightly deviate from 1. %
The first order phase transition at finite T for model (III) with a confinement-deconfinement phase transition occurs because the gap is not a smooth function of $m_d$.

The most important conceptual problem is the exponential growing of meson polarization loops in Minkowski space.
This issue arises because if the quark propagator has no singularities at finite $p^2$, it will have an essential singularity at infinity.
How to solve this problem remains unclear, but sub-leading $1/N_c$ corrections, such as pion dressing of the
quark propagator, may introduce physically motivated singularities.
The pion-quark dressing is discussed by V. N. Gribov in his confinement theory \cite{Gribov:1999ui}.

This work is supported by supported by the CAS President’s international fellowship initiative (Grant No. 2023VMA0015); the project of the Ministry of Education and Science of the Russian Federation (”Analytical and numerical methods of mathematical physics in problems of tomography,quantum field theory, fluid and gas mechanics”no. 121041300058-1); the National Natural Science Foundation of China under Grants No. 12375117; and the Youth Innovation Promotion Association of Chinese Academy of Sciences (Grant No. Y2021414).

The authors thank the I.V. Anikin, Yu.M. Bystritskiy, V.P. Lomov and B. Zhang for fruitful comments.  %

\end{document}